\documentclass[11pt]{article}
\usepackage[dvips]{graphicx}
\usepackage{mathrsfs}
\usepackage{amsthm} 
\usepackage{amsmath}
\usepackage{amssymb}
\usepackage{amsfonts}
\usepackage{bbm}
\usepackage{float}
\usepackage{yhmath}
\usepackage{wrapfig}
\usepackage[dvips]{color}
\usepackage[dvips]{colortbl}
\usepackage{cite}
\usepackage{array}
\usepackage{tabularx}
\usepackage[sectionbib]{chapterbib}
\usepackage{threeparttable}
\usepackage{chemarrow}
\usepackage{framed}
\usepackage{ascmac}
\definecolor{shadecolor}{gray}{0.90}
\usepackage[small, bf, labelsep=quad]{caption}
\usepackage{cases}
\newcommand{\II}{I\hspace{-0.5mm}I}

\definecolor{mycolor1}{rgb}{1,1,0.7}
\definecolor{mycolor2}{rgb}{0.9,1,1}
\definecolor{mycolor3}{cmyk}{0,0,0,0.113}
\definecolor{mycolor4}{cmyk}{0.086,0,0,0}

\begin{document}

\begin{center}
\Huge{Segment Distribution around the Center of Gravity of Ring Polymers}
\end{center}
\vspace*{-3mm}
\begin{center}
\LARGE{Extension of the Kramers Method}
\end{center}

\vspace*{0mm}
\begin{center}
\large{Kazumi Suematsu\footnote{\, The author takes full responsibility for this article.}, Haruo Ogura$^{2}$, Seiichi Inayama$^{3}$, and Toshihiko Okamoto$^{4}$} \vspace*{2mm}\\
\normalsize{\setlength{\baselineskip}{12pt} 
$^{1}$ Institute of Mathematical Science\\
Ohkadai 2-31-9, Yokkaichi, Mie 512-1216, JAPAN\\
E-Mail: suematsu@m3.cty-net.ne.jp, ksuematsu@icloud.com  Tel/Fax: +81 (0) 593 26 8052}\\[3mm]
$^{2}$ Kitasato University,\,\, $^{3}$ Keio University,\,\, $^{4}$ Tokyo University\\[12mm]
\end{center}

%%%%%%%%%%%%%%%%%%
\hrule
\vspace{3mm}
\noindent
\textbf{\large Abstract}: 
The segment distribution around the center of gravity is derived for unperturbed ring polymers. We show that, although a small difference is observed, the exact distribution can be well approximated by the Gaussian probability distribution function.
\vspace{0mm}
\begin{flushleft}
\textbf{\textbf{Key Words}}: Unperturbed Ring Polymers/ Segment Distribution/
\normalsize{}\\[3mm]
\end{flushleft}
\hrule
\vspace{5mm}
\setlength{\baselineskip}{14pt}

\vspace*{0mm}
\section{Derivation}
The mean radius of gyration of a simple ring has already been derived\cite{Zimm, Kramers}. In this paper, we derive the segment probability distribution around the center of gravity of an unperturbed ring polymer, making an extension of the Kramers method\cite{Kramers, Zimm}.

\begin{figure}[h]
\vspace{0mm}
\begin{center}
\includegraphics[width=11cm]{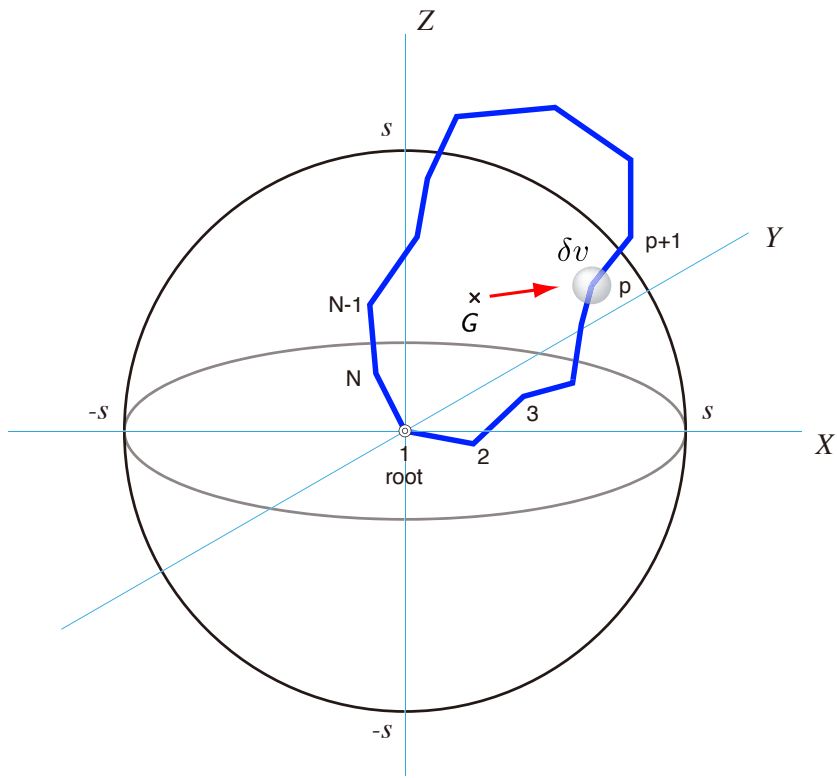}
\caption{The 3D view of a ring molecule having $N$ monomers.}\label{RingMolecule3D}
\end{center}
\end{figure}

\begin{figure}[h]
\begin{center}
\includegraphics[width=16cm]{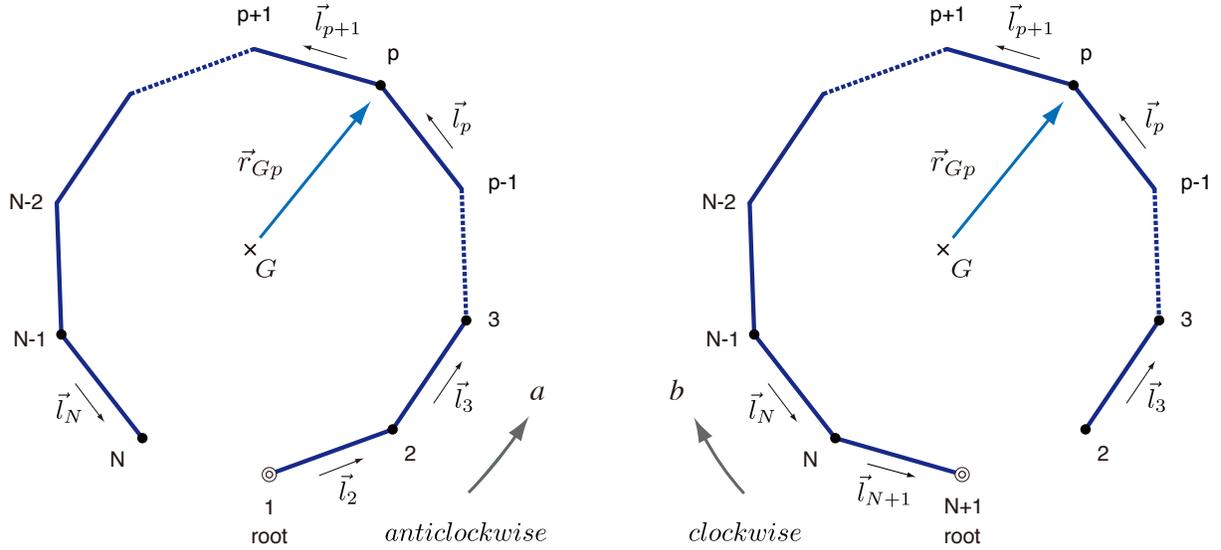}
\caption{The two perspectives of a ring molecule having $N$ monomers. \textit{G} ($\times$) represents the center of masses, and 1 ($\equiv$ N+1) the root monomer.}\label{RingA}
\end{center}
\vspace*{0mm}
\end{figure}

Consider an $N$-ring comprised of $N$ monomers. Let an arbitrary monomer on the ring be the root (the symbol $\circledcirc$ in Figs. \ref{RingMolecule3D} and \ref{RingA}). We consider the vector, $\vec{r}_{Gp}$, from the center of masses to an arbitrary monomer $p$. As usual, we use the Isihara formula\cite{Isihara}:
%%%%%%%%%%%%%%%%%% Eq. 1
\begin{equation}
\vec{r}_{Gp}=\vec{r}_{1p}-\frac{1}{N}\sum_{p=1}^{N}\vec{r}_{1p}\label{Ring-1}
\end{equation}
Let us index the monomers, in the anticlockwise direction $a$, from 1 (root), 2, 3, $\cdots$, $N$; or equivalently, in the clockwise direction $b$, from $N+1$ ($\equiv$1), $N$, $N-1$, $\cdots$, 2 (see Figs. \ref{RingMolecule3D} and \ref{RingA}). Consider the two chain molecules of 1 to $N$ (anticlockwise) and $N+1$ to 2 (clockwise). What we should do is only to decompose the vector, $\vec{r}_{Gp}$, into the grand sum of all bond vectors $\{\vec{l}_{k}\}$. We define the bond vectors, in the anticlockwise direction, such that $\vec{l}_{k}=\vec{r}_{k}-\vec{r}_{k-1}$. Let all the bonds have the same length: $|\vec{l}_{k}|=l$. Applying Eq. (\ref{Ring-1}) to the two chains that extend oppositely to each other (Fig. \ref{RingA}), we have the expressions of $\vec{r}_{Gp}$ as functions of $p$:
\begin{description}
\item[(1)] along the path $a$
%%%%%%%%%%%%%%%%%% Eq. 2
\begin{equation}
\vec{r}_{Gp}(a)=\frac{1}{N}\left\{\sum_{k=1}^{p-1}k\,\,\vec{l}_{k+1}-\sum_{k=p}^{N-1}(N-k)\vec{l}_{k+1}\right\}\label{Ring-2}
\end{equation}
\item[(2)] along the path $b$
%%%%%%%%%%%%%%%%%% Eq. 3
\begin{equation}
\vec{r}_{Gp}(b)=\frac{1}{N}\left\{\sum_{k=2}^{p-1}(k-1)\vec{l}_{k+1}-\sum_{k=p}^{N}\left[N-k+1\right]\vec{l}_{k+1}\right\}\label{Ring-3}
\end{equation}
\end{description}
The use of the notational identity, $\vec{l}_{N+1}\equiv\vec{l}_{1}$, is simply for mathematical convenience. 

The respective vectors of Eqs. (\ref{Ring-2}) and (\ref{Ring-3}) are expressions for the typical end-to-end distances from the center of masses to the $p$th monomer, with unequal step lengths\cite{Weiss, Redner, Kazumi2}, which can be recast into the grand sum of all bond vectors that constitute the ring. So, the vector, $\vec{r}_{Gp}(a)$, represents an $N$ chain having $N-1$ bonds joined by unequal step lengths from the root monomer 1 to the $N$th monomer anticlockwisely, and the vector, $\vec{r}_{Gp}(b)$, represents the corresponding $N$ chain having $N-1$ bonds from the $(N+1)$th monomer to the 2nd monomer clockwisely:
%%%%%%%%%%%%%%%%%% Eq. 4
\begin{align}
\vec{r}_{Gp}(a)&=\frac{1}{N}\sum_{j=2}^{N}c_{j}(p)\,\vec{l}_{j}\label{Ring-4}\\
\vec{r}_{Gp}(b)&=\frac{1}{N}\sum_{j=3}^{N+1}c_{j}'(p)\,\vec{l}_{j}\label{Ring-5}
\end{align}
A necessary and sufficient condition for the polymer to be a closed ring is
%%%%%%%%%%%%%%%%%% Eq. 6
\begin{equation}
\vec{r}_{Gp}(a)\equiv\vec{r}_{Gp}(b)\label{Ring-6}
\end{equation}
which, by Eqs. (\ref{Ring-2}) and (\ref{Ring-3}), leads us to
%%%%%%%%%%%%%%%%%% Eq. 7
\begin{equation}
\sum_{k=1}^{N}\vec{l}_{k+1}=\vec{0}\label{Ring-7}
\end{equation}
namely 
%%%%%%%%%%%%%%%%%% Eq. 8
\begin{equation}
\vec{l}_{N+1}=-\sum_{k=1}^{N-1}\vec{l}_{k+1}\label{Ring-8}
\end{equation}
Upon substituting Eq. (\ref{Ring-8}) back into Eq. (\ref{Ring-3}), we recover the mathematical identity (\ref{Ring-6}). Eq. (\ref{Ring-7}) is a different expression of the necessary and sufficient condition.

The respective vectors of Eqs. (\ref{Ring-2}) and (\ref{Ring-3}) have the forms of $N$ chains represented by Eqs. (\ref{Ring-4}) and (\ref{Ring-5}), so that those, given the random walk assumption, should approach, as $N\rightarrow\infty$, the Gaussian distribution (PDF) having the mean squares of the end-to-end distances of the forms:
\begin{description}
\item[(1)] for the path $a$
%%%%%%%%%%%%%%%%%% Eq. 9
\begin{equation}
\left\langle{r}_{Gp}^{\hspace{0.5mm}2}(a)\right\rangle=\frac{1}{N^{2}}\left\{\sum_{k=1}^{p-1}k^{2}+\sum_{k=p}^{N-1}(N-k)^{2}\right\}l^{2}\label{Ring-9}
\end{equation}
\item[(2)] for the path $b$
%%%%%%%%%%%%%%%%%% Eq. 10
\begin{equation}
\left\langle{r}_{Gp}^{\hspace{0.5mm}2}(b)\right\rangle\equiv\left\langle{r}_{Gp}^{\hspace{0.5mm}2}(a)\right\rangle\label{Ring-10}
\end{equation}
\end{description}
By the definition of the ring molecule, the $p$th monomer as the end monomer on the vector, $\vec{r}_{Gp}$, must occupy the same position in the volume element, $\delta V=S_{d}(s)\delta s$, for both the paths, $a$ and $b$ $\left[\right.$$S_{d}(s)$ is the surface area of the $d$-dimensional sphere$\left.\right]$. This condition is fulfilled by imposing the requirement that the $p$th monomer on the vector, $\vec{r}_{Gp}(a)$, must lie within the small volume element, $\delta v$, around the $p$th monomer on the vector, $\vec{r}_{Gp}(b)$, the position of which oscillates thermally around the average coordinates (Fig. \ref{RingMolecule3D}). Since the locus of $\vec{r}_{Gp}$ is Gaussian when $N$ is large\cite{Kazumi2}, the probability of the end-to-end vector lying between $s$ and $s+\delta s$ should satisfy
%%%%%%%%%%%%%%%%%%
\begin{equation}
\varphi_{\textsf{\textit{G}}p}\,S_{d}(s)\delta s=const.\exp\left(-\frac{d}{2\left\langle{r}_{Gp}^{\hspace{0.5mm}2}(a)\right\rangle}\,s^{2}\right)S_{d}(s)\delta s\cdot\exp\left(-\frac{d}{2\left\langle{r}_{Gp}^{\hspace{0.5mm}2}(b)\right\rangle}\,s^{2}\right)S_{d}(s)\delta s\cdot\left(\delta v/\delta V\right)\notag
\end{equation}
Substituting Eq. (\ref{Ring-10}), the above equation can be recast in the form:
%%%%%%%%%%%%%%%%%% Eq. 11
\begin{equation}
\varphi_{\textsf{\textit{G}}p}\,S_{d}(s)\delta s=const.\left[\exp\left(-\frac{d}{2\left\langle{r}_{Gp}^{\hspace{0.5mm}2}(a)\right\rangle}\,s^{2}\right)S_{d}(s)\delta s\right]^{2}\left(\delta v/\delta V\right)\label{Ring-11}
\end{equation}
The small volume element, $\delta v$, may be assumed to be independent of the end-to-end distance, $\vec{r}_{Gp}$, so that it can be absorbed into the normalization constant, $\mathscr{N}$, in conjunction with the constant term. Substituting the equality, $\delta V=S_{d}(s)\delta s$, into Eq. (\ref{Ring-11}), we have finally
%%%%%%%%%%%%%%%%%% Eq. 12
\begin{equation}
\varphi_{\textsf{\textit{G}}p}\,S_{d}(s)\delta s=\mathscr{N}\cdot\exp\left(-\frac{d}{2\left\langle{r}_{\textsf{\textit{G}}p}^{\hspace{0.5mm}2}\right\rangle}\,s^{2}\right)S_{d}(s)\delta s\label{Ring-12}
\end{equation}
with $\mathscr{N}=const.\,\delta v$. Comparing Eq. (\ref{Ring-11}) and (\ref{Ring-12}), one has
%%%%%%%%%%%%%%%%%% Eq. 13
\begin{equation}
\left\langle{r}_{\textsf{\textit{G}}p}^{\hspace{0.5mm}2}\right\rangle=\frac{1}{2}\left\langle{r}_{Gp}^{\hspace{0.5mm}2}(a)\right\rangle \hspace{3mm}(p=1, 2, \cdots, N)\label{Ring-13}
\end{equation}

%%%%%%%%%%%%%%%%%% Fig. 1
%%%%%%%%%%%%%%%%%% Fig. 2

Since all monomers are joined by the chemical bonds, the configurational trajectory of each monomer can not be independent. 
Hence, the spatial distribution of the monomers on the ring molecule should be expressed by the average of the grand sum of each trajectory:
%%%%%%%%%%%%%%%%%% Eq. 14
\begin{equation}
\varphi_{{ring}}(s)=\frac{1}{N}\sum_{p=1}^{N}\left(\frac{d}{2\pi\left\langle{r}_{\textsf{\textit{G}}p}^{\hspace{0.5mm}2}\right\rangle}\right)^{d/2}\exp\left(-\frac{d}{2\left\langle{r}_{\textsf{\textit{G}}p}^{\hspace{0.5mm}2}\right\rangle}\,s^{2}\right)\label{Ring-14}
\end{equation}
The mean square radius of gyration can be evaluated from $\varphi_{{ring}}(s)$ by the equation:
%%%%%%%%%%%%%%%%%% Eq. 15
\begin{equation}
\left\langle{s}_{N}^{\hspace{1mm}2}\right\rangle_{ring}=\int_{0}^{\infty}s^{2}\varphi_{{ring}}(s)S_{d}(s)ds=\frac{1}{N}\sum_{p=1}^{N}\left\langle{r}_{\textsf{\textit{G}}p}^{\hspace{0.5mm}2}\right\rangle\label{Ring-15}
\end{equation}
which, by Eqs. (\ref{Ring-9}), (\ref{Ring-13}), and (\ref{Ring-15}), yields immediately
%%%%%%%%%%%%%%%%%% Eq. 16
\begin{equation}
\left\langle{s}_{N}^{\hspace{1mm}2}\right\rangle_{ring}=\frac{1}{12}\left(N-\frac{1}{N}\right)l^{2}\label{Ring-16}
\end{equation}
For a large $N$, Eq. (\ref{Ring-16}) recovers the Kramers result\cite{Kramers, Zimm, Fujita}:
%%%%%%%%%%%%%%%%%% Eq. 17
\begin{equation}
\left\langle{s}_{N}^{\hspace{1mm}2}\right\rangle_{ring}=\frac{1}{12}N\hspace{0.3mm}l^{2}\hspace{5mm}(N\rightarrow\infty) \label{Ring-17}
\end{equation}

%%%%%%%%%%%%%%%%%% Fig. 3
\begin{figure}[H]
\vspace*{0mm}
\begin{center}
\includegraphics[width=8cm]{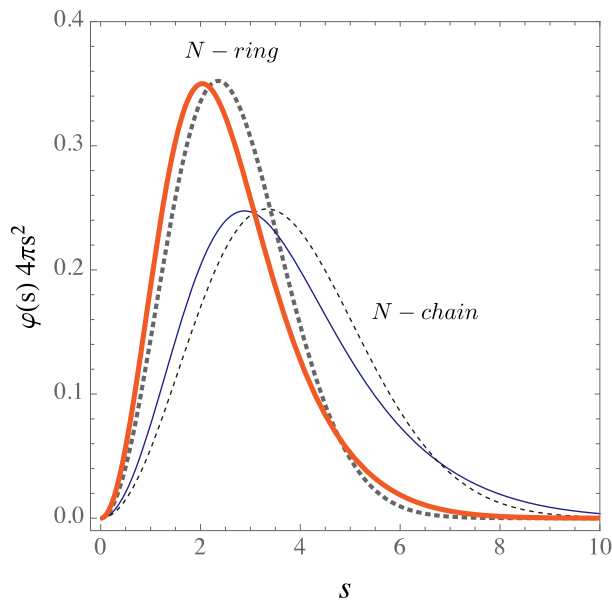}
\caption{The radial mass distribution around the center of gravity of an unperturbed ring polymer having $N=100$ monomers. The thick solid line: the exact result calculated by Eq. (\ref{Ring-14}); the thick dotted line: the Gaussian function having the mean square of the radius of gyration represented by Eq. (\ref{Ring-17}). The thin solid line: the corresponding distribution of an $N-$chain; the thin dotted line: the Gaussian approximation for the $N-$chain.}\label{SD-Ring}
\end{center}
\end{figure}

In Fig. \ref{SD-Ring}, we plot Eq. (\ref{Ring-14}) with the help of Eqs. (\ref{Ring-9}) and (\ref{Ring-13}). The thick solid line in Fig. \ref{SD-Ring} represents the exact result according to Eq. (\ref{Ring-14}), and the thick dotted line the Gaussian PDF having the mean square of the radius of Eq. (\ref{Ring-17}). It is seen that while there is a small deviation from the exact result, the Gaussian PDF is a good approximation of Eq. (\ref{Ring-14}). The corresponding mass distribution (thin solid line) around the center of gravity for an $N-$chain and its Gaussian approximation (thin dotted line) are shown for comparison. As expected, the masses on the $N-$ring distribute nearer the center of gravity than those on the $N-$chain.

%\vspace{10mm}
%%%%%%%%%%%%%%%%%%

\end{document}